\documentclass[a4paper,12pt]{amsart}

\usepackage{amsfonts}
\usepackage{amsmath}
\usepackage{amssymb}
\usepackage{mathrsfs}
\usepackage{hyperref}
\usepackage{graphicx}
\usepackage{algorithm}
\usepackage{listings}
\usepackage[noend]{algorithmic}
\usepackage{amsmath,tikz}
\usetikzlibrary{matrix}

\numberwithin{equation}{section}
\theoremstyle{plain}

\theoremstyle{definition}

\begin{document}

\title[A parallel hybrid implementation of the 2D acoustic wave equation]
{A parallel hybrid implementation of the 2D acoustic wave equation}

\author[Arshyn Altybay]{Arshyn Altybay}
\address{
  Arshyn Altybay:
  \endgraf
  Al-Farabi Kazakh National University
  \endgraf
  71 Al-Farabi avenue
  \endgraf
  050040 Almaty
  \endgraf
  Kazakhstan
  \endgraf
  and
  \endgraf
  Department of Mathematics: Analysis, Logic and Discrete Mathematics
  \endgraf
  Ghent University, Belgium
  \endgraf
  and
  \endgraf
  Institute of Mathematics and Mathematical Modeling
  \endgraf
  125 Pushkin str., Almaty, 050010
  \endgraf
  Kazakhstan,
  \endgraf
{\it E-mail address} {\rm arshyn.altybay@gmail.com}}

\author[Michael Ruzhansky]{Michael Ruzhansky}
\address{
  Michael Ruzhansky:
  \endgraf
  Department of Mathematics: Analysis, Logic and Discrete Mathematics
  \endgraf
  Ghent University, Belgium
  \endgraf
  and
  \endgraf
  School of Mathematical Sciences
  \endgraf
  Queen Mary University of London
  \endgraf
  United Kingdom
  \endgraf
{\it E-mail address} {\rm michael.ruzhansky@ugent.be}}

\author[Niyaz Tokmagambetov]{Niyaz Tokmagambetov}
\address{
  Niyaz Tokmagambetov:
  \endgraf
  Department of Mathematics: Analysis, Logic and Discrete Mathematics
  \endgraf
  Ghent University, Belgium
  \endgraf
  and
  \endgraf
  al--Farabi Kazakh National University
  \endgraf
  71 al--Farabi ave., Almaty, 050040
  \endgraf
  Kazakhstan,
  \endgraf
{\it E-mail address} {\rm niyaz.tokmagambetov@ugent.be}}

\thanks{The authors were supported  by the FWO Odysseus 1 grant G.0H94.18N: Analysis and Partial Differential Equations. MR was supported in parts by the EPSRC Grant EP/R003025/1, by the Leverhulme Research Grant RPG-2017-151. AA was supported by the MESRK Grants AP08052028 and AP08053051 of the Ministry of Education and Science of the Republic of Kazakhstan}



\begin{abstract}
In this paper, we propose a hybrid parallel programming approach for a numerical solution of a two-dimensional acoustic wave equation using an implicit difference scheme for a single computer.
The calculations are carried out in an implicit finite difference scheme.
First, we transform the differential equation into an implicit finite-difference equation and then using the ADI method, we split the equation into two sub-equations.
Using the cyclic reduction algorithm, we calculate an approximate solution. Finally, we change this algorithm to parallelize on GPU, GPU+OpenMP, and Hybrid (GPU+OpenMP+MPI) computing platforms.

The special focus is on improving the performance of the parallel algorithms to calculate the acceleration based on the execution time. We show that the code that runs on the hybrid approach gives the expected results by comparing our results to those obtained by running the same simulation on a classical processor core, CUDA, and CUDA+OpenMP implementations.

\end{abstract}

\maketitle
\section{Introduction}

The reduction of computational time for long-term simulation of physical processes is a challenge and an important issue in the field of modern scientific computing.
The cost of supercomputer, CPU clusters and hybrid clusters with a large number of GPUs are very expensive and they consume a lot of energy, which is inaccessible and ineffective to some small laboratories and individuals.

Nowadays, new generation computers are multi-core, hybrid architecture and their computational power is also quite high. For example, the Intel Xeon E5-2697 v2 (2S-E5) processors theoretically has computing power of about 19.56 GFLOPS, and, accordingly, the computational power of the NVIDIA TITAN Xp video card is about up to 379.7 GFLOPS. If we use the computing power of the CPU and GPU  together, we can show good results.

The goal of this work is to develop a parallel hybrid implementation of the finite-difference method for solving two-dimensional wave equation using CUDA, CUDA + OpenMP and CUDA + OpenMP + MPI technologies and to study the parallelization efficiency by comparing the time to solve this problem with the above approaches.
Already for several years, GPUs have been used to accelerate well parallelizable computing, only with the advent of a new generation of GPUs with multicore architecture, this direction began to give palpable results.

For multidimensional problems, the efficiency of an implicit compact difference scheme depends on the computational efficiency of the corresponding matrix solvers. From this point of view, the ADI method \cite{PR} is promising because they can decompose a multidimensional problem into a series of one-dimensional problems. It has been shown that schemes acquired are unconditionally stable.
For the proper assignment of large domains of modeling, two- or three-dimensional computational grids with a sufficient number of points are used. Calculations on such grids require more CPU time and computer memory resources. To accelerate the computation process, GPU, OpenMP, MPI technologies were used in this paper, which allows the program to operate on larger grids.
With GPU becoming a viable alternative to CPU for parallel computing, the aforementioned parallel tridiagonal solvers and other hybrid methods have been implemented on GPUs \cite{Zhang10}--\cite{Wei}.
In this paper, we propose three different parallel programming approaches using hybrid CUDA, OpenMP and MPI programming for personal computers.
There are many examples in the literature of successfully using hybrid approaches for different simulation \cite{Fra}--\cite{Alonso}.

Here we study some issues in the numerical simulation of some problems in the propagation of acoustic waves on high performance computing systems.

\section{Problem Statement and Numerical Scheme}

We consider 2D acoustic wave equation with the positive "speed" function $c$ and the source term $f$
\begin{equation}\label{f1}
\frac{\partial^2 u}{\partial t^2}-c(t)\left(\frac{\partial^2 u}{\partial x^2}+\frac{\partial^2 u}{\partial y^2}\right)=f(t, x, y), \,\, (t, x, y)\in [0;T]\times [0;l]\times [0;l],
\end{equation}
subject to the initial conditions
\begin{equation}\label{f2}
u(0,x,y) = \varphi (x,y), \,\, x,y\in[0,l],
\end{equation}
\begin{equation}\label{f3}
\frac{\partial u(0,x,y)}{\partial t} = \psi (x,y), \,\, x,y\in[0,l],
\end{equation}
and boundary conditions
\begin{equation}\label{f4}
u(t,x,0)=0, \,\, u(t,x,l)=0, \,\, t\in[0,T], \,\, x\in[0,l],
\end{equation}
\begin{equation}\label{f5}
u(t,0,y)=0, \,\, u(t,l,y)=0, \,\, t\in[0,T], \,\, y\in[0,l].
\end{equation}
In what follows, we take all data, namely, the coefficient $c$, the source function $f$, the initial functions $\varphi$ and $\psi$, smooth enough.  Due to the notion of "very weak solutions" and the approach developed by Garetto and Ruzhansky in \cite{GR15}, we can deal with 2D acoustic wave equation with singular ($\delta$--like) data approximating them by smooth functions. For more details on these techniques and applications, we refer to the papers \cite{GR15, RT17a, RT17b, RT18, MRT19a, MRT19b, RT19}.

For numerical simulations, we introduce a space-time grid with steps $h_1, h_2, \tau$ respectively, in the variables $x, y, t:$
\begin{equation}\label{f6}
\omega_{h_1,h_2}^\tau={\{t_k=k\tau, k=\overline{1, M}; \,\, x_i=ih_1,i=\overline{1, N_1}; \,\, y_j=jh_2,j=\overline{1, N_2}\}},
\end{equation}
and
\begin{equation}\label{f6}
\Omega_{h_1,h_2}^\tau={\{t_k=k\tau, k=\overline{0,M}; \,\, x_i=ih_1,i=\overline{0,N_1}; \,\, y_j=jh_2,j=\overline{0,N_2}\}},
\end{equation}
where $h_1=l/N_1$, $h_2=l/N_2$ and $\tau=T/M$.

On this grid we approximate the problem \eqref{f1}--\eqref{f5} using the finite difference method. For simplicity, we put $N:=N_1=N_2$ and denote $h:=h_1=h_2$. Consider the Crank-Nicolson scheme for the problem \eqref{f1}--\eqref{f5}
\begin{equation}\label{CN1}
\begin{split}
\frac{u_{i,j}^{k+1}-2u_{i,j}^k+u_{i,j}^{k-1}}{\tau^2}&-\frac{c^{k+1}}{2h^2}(u_{i+1,j}^{k+1}
-2u_{i,j}^{k+1}+u_{i-1,j}^{k+1}+u_{i,j+1}^{k+1}
-2u_{i,j}^{k+1}+u_{i,j-1}^{k+1}) \\ -&\frac{c^{k-1}}{2h^2}(u_{i+1,j}^{k-1}-2u_{i,j}^{k-1}+u_{i-1,j}^{k-1}+u_{i,j+1}^{k-1}-2u_{i,j}^{k-1}+u_{i,j-1}^{k-1})
= f_{i,j}^k,
\end{split}
\end{equation}
for $(k, i, j) \in \omega_{h_1,h_2}^\tau,$ with initial conditions
\begin{equation}\label{CN2}
\begin{split}
u_{i,j}^{0}= \varphi_{i,j}, \,\,\, u_{i,j}^{1} - u_{i,j}^{0}= \tau \psi_{i,j},
\end{split}
\end{equation}
for $(i, j) \in  \overline{0, N}\times\overline{0, N},$ and with boundary conditions
\begin{equation}\label{CN3}
\begin{split}
u_{0, j}^{k}= 0, \,\,\, u_{N, j}^{k}= 0, \,\,\, u_{i, 0}^{k}= 0, \,\,\, u_{i, N}^{k}= 0, \,\,\,
\end{split}
\end{equation}
for $(j, k) \in  \overline{0, N}\times\overline{0, M}$ and $(i, k) \in  \overline{0, N}\times\overline{0, M}$, respectively.

It is well-known, that the implicit scheme is unconditionally stable and it has accuracy order $O (\tau+|h|^2 )$, see, for example \cite{Sam01}. We solve the difference equation \eqref{CN1} by the alternating direction implicit (ADI) method, namely, dividing it into two sub-problems
\begin{equation}\label{fr1}
\begin{split}
&\frac{u_{i,j}^{k+1/2}-2u_{i,j}^k+u_{i,j}^{k-1/2}}{\tau^2}-\frac{c^{k+1/2}}{2h^2}(u_{i+1,j}^{k+1/2}-2u_{i,j}^{k+1/2}+u_{i-1,j}^{k+1/2})\\
&-\frac{c^{k-1/2}}{2h^2}(u_{i+1,j}^{k-1/2}-2u_{i,j}^{k-1/2}+u_{i-1,j}^{k-1/2})
= f_{i,j}^k,
\end{split}
\end{equation}
and
\begin{equation}\label{fr2}
\begin{split}
&\frac{u_{i,j}^{k+1}-2u_{i,j}^{k+1/2}+u_{i,j}^k}{\tau^2}-\frac{c^{k+1}}{2h^2}(u_{i,j+1}^{k+1}
-2u_{i,j}^{k+1}+u_{i,j-1}^{k+1})\\
&-\frac{c^{k}}{2h^2}(u_{i,j+1}^{k}-2u_{i,j}^{k}+u_{i,j-1}^{k}) = f_{i,j}^{k+1/2}.
\end{split}
\end{equation}

\section {Hybrid parallel computing model}
High-performance computing uses parallel computing to achieve high levels of performance. In parallel computing, the program is divided into many subroutines, and then they are all executed in parallel to calculate the required values.
In this section, we will propose a hybrid parallel approach numerically solving a two-dimensional wave equation, for this, we use CUDA, MPI OpenMP technologies.

\subsection{CUDA approach}
The graphics processing unit (GPU) is a highly parallel, multi-threaded, and multi-core processor with enormous processing power. Its low cost and high bandwidth floating point operations and memory access bandwidth are attracting more and more high performance computing researchers \cite{KWBH}. In addition, compared to cluster systems, which consist of several processors, computing on a GPU is inexpensive and requires low power consumption with equivalent performance. In many disciplines of science and technology, users were able to increase productivity by several orders of magnitude using graphics processors \cite{BG}, \cite{ELD}.
The year 2007, with the appearance of the CUDA programming language, programming GPUs on NVIDIA graphics cards became significantly simpler because its syntax is similar to C\cite{NVI}.

It is designed so that its constructions allow a natural expression of concurrency at the data level. A CUDA program consists of two parts: a sequential program running on the CPU, and a parallel part running on the GPU \cite{ELD}, \cite{NBGS}. The parallel part is called the kernel. A CUDA program automatically uses more parallelism on GPUs that have more processor cores.

A C program using CUDA extensions hand out a large number of copies of the kernel into available multiprocessors to be performed simultaneously.

The CUDA code consists of three computational steps: transferring data to the global GPU memory, running the CUDA core, and transferring the results from the GPU to the CPU memory.
We have designed a CUDA program based on cyclic reduction  method, whose full CR function codes are located in  \cite{pr_c}. The algorithm for solving the problem \eqref{f1}--\eqref{f5} is shown in Algorithm \ref{alg:the_alg}.
\begin{algorithm}
\caption{Implementation of 2D wave equation }\label{alg:the_alg}
\begin{algorithmic}
\STATE compute initial function matrix $U0$
\STATE from initial condition \eqref{f2} we get $u=U0$
\WHILE{($t<t_{end}$)}
\STATE for $j=0,\dots,n$ \par
       for $i=0,\dots,n$ \par
   calculate tridiagonal system elements $a_i,b_i,c_i,f_i$ \par
   call function $CR(a_i,b_i,c_i,f_i,y_i,n)$ \par
   calculate matrix $Ux$ \par
\STATE for $i=0,\dots,n$ \par
       for $j=0,\dots,n$ \par
   calculate tridiagonal system elements $aj,bj,cj,fj$ \par
   call function $CR(a_j,b_j,c_j,f_j,y_j,n)$ \par
   calculate matrix $Uy$
\STATE swap $(u,Ux)$ \par
       swap $(U0,Uy)$
\STATE $t \leftarrow t+\vartriangle t$
\ENDWHILE
\end{algorithmic}
\end{algorithm}

Here, $u$, $U0$, $Ux$, $Uy$ denote $u_{i,j}^{k-1/2}$,  $u_{i,j}^{k}$, $u_{i,j}^{k+1/2}$, $u_{i,j}^{k+1}$, respectively. The $CR()$ function includes 3 device functions, namely,  $CRM\_forward(), cr\_div(),\\ and CRM\_backward()$, and one host function $calc\_dim()$. First we have to calculate the block size according to the size of the matrix and step numbers of forward and backward sub-steps. For this, we use one cycle \par
	  for $(i = 0; i < \log 2(n + 1) - 1; i++)$ $\{$ \par
	    $stepNum = (n - pow(2.0, i + 1)) / pow(2.0, i + 1) + 1;$ \par
		$calc\_dim(stepNum, \&dimBlock, \&dimGrid)$; \par
		$CRM\_forward$ $<<< dimGrid, dimBlock  >>>$($d\_a, d\_b, d\_c, d\_f, n, stepNum, i$); \par
	$\}$\par
Here $\log 2(n + 1)-1$ is a step number and a variable of $stepNum$. It is to identify the block size. Therefore, we need the function $calc\_dim()$, which is identifying the block sizes. After that the function $CRM\_forward()$ runs $\log 2(n + 1)-1$ times. Consequently, the system reduces to one equation.
After that we synchronize the device and call the function $cr\_div()$. This function calculates two unknowns. Then we use one cycle \par
for ($i = \log 2(n + 1) - 2; i>$ = $0; i$ $-$ $-$) $\{$ \par
$stepNum = (n - pow(2.0, i + 1)) / pow(2.0, i + 1) + 1;$ \par
$calc\_dim(stepNum, \&dimBlock, \&dimGrid);$\par
$CRM\_backward$ $<<<dimGrid, dimBlock >>>$($d\_a, d\_b, d\_c, d\_f, d\_x, n, stepNum, i$);\par
	$\}$\par
Here the backward substitution runs $\log 2(n + 1) - 2$ times because the first backward substitution sub-step is calculated by the function $calc\_dim()$. Thus, we calculate $d\_x$ array. After that
we copy the calculated data $d\_x$ from the device to the host using \par
$cudaMemcpy(y, $d\_x$, sizeof(double)*n, cudaMemcpyDeviceToHost)$.

\subsection{OpenMP+CUDA approach}

OpenMP (Open Multi-Processing) was introduced to provide the means to implement shared memory concurrency in FORTRAN and C/C ++ programs. In particular, OpenMP defines a set of environment variables, compiler directives and library procedures that will be used for parallelization with shared memory. OpenMP was specifically designed to use certain characteristics of shared memory architectures, such as the ability to directly access memory throughout a system with low latency and very fast shared memory locking \cite{Geo}.

We can easy parallelize loops by using MPI thread libraries and invlove the OpenMP compilers. In this way, the threads can obtain new tasks, the unprocessed loop iterations directly from local shared memory. The basic idea behind OpenMP is data-shared parallel execution. It consists of a set of compiler directives, callable runtime library routines and environment variables that extend FORTRAN, C, and C++ programs. The working unit of OpenMP is a thread. It works well when accessing shared data costs you nothing. Every thread can access a variable in a shared cache or RAM.

In this paper, we use OpenMP to solve an initial boundary value problem. Since do deal with it we use two cycles and calculate one matrix. Moreover, OpenMP parallel computing model is very convenient to implement, and it has low latency and high bandwidth.

\subsection{Hybrid approach}
The message passing interface (MPI) is a standardized and portable programming interface for exchanging messages between multiple processors executing a parallel program in distributed memory.

MPI works well on a wide variety of distributed storage architectures and is ideal for individual computers and clusters. However, MPI depends on explicit communication between parallel processes which requires mesh decomposition in advance due to data decomposition.  Therefore, MPI can cause load balancing and consume extra time.

Since MPICH2 is freely accessible here in our implementations we use it. In  \cite{Goddeke} the authors used a compact implementation of the MPI standard for message-passing for distributed-memory applications. MPICH is a free software. Also, it is available for most types of UNIX and Microsoft Windows systems.
MPI is standardized on many levels. Indeed, it provides many advantages for the users. One of them makes you sure that MPI codes can be executed in any MPI implementation launching on your architecture, even if the syntax has been standardized.
Since the functional behavior of the MPI calls is also standardized, its should behave in the same way whatever of implementation, which ensures the portability of the parallel programs.

We use MPI technology to calculate the elements of the tridiagonal matrix system, i.e $ai, bi, ci, fi$ because these values can be calculated independently, so we can successfully apply MPI technology here.

Listing code \ref{list1} shows the program code.
\begin{lstlisting}[caption={Calculation of $ai, bi, ci, fi$}, label={list1}, language=c]

	i1 = (n*rank) / size;
	i2 = (n*(rank + 1)) / size;
for (i = i1; i <i2; i++)
		{
			a_m[kk] = tau*tau;
			c_m[kk] = tau*tau;
			b_m[kk] = 2 * tau*tau + h*h;
			f_m[kk] = h*h*Unn[i] - 2 * h*h*uu0[i];
			kk++;
		}

		MPI_Gather(a_m, n/size, MPI_DOUBLE, a, n/size, MPI_DOUBLE, 0, MPI_COMM_WORLD);
		MPI_Gather(b_m, n/size, MPI_DOUBLE, b, n/size, MPI_DOUBLE, 0, MPI_COMM_WORLD);
		MPI_Gather(c_m, n/size, MPI_DOUBLE, c, n/size, MPI_DOUBLE, 0, MPI_COMM_WORLD);
		MPI_Gather(f_m, n/size, MPI_DOUBLE, f, n/size, MPI_DOUBLE, 0, MPI_COMM_WORLD);
		
		if (rank == 0)
		{
			MPI_Bcast(a, n, MPI_DOUBLE, 0, MPI_COMM_WORLD);
			MPI_Bcast(b, n, MPI_DOUBLE, 0, MPI_COMM_WORLD);
			MPI_Bcast(c, n, MPI_DOUBLE, 0, MPI_COMM_WORLD);
			MPI_Bcast(f, n, MPI_DOUBLE, 0, MPI_COMM_WORLD);
		}
\end{lstlisting}

\begin{figure}[ht!]
\begin{minipage}[h]{0.75\linewidth}
\center{\includegraphics[scale=0.61]{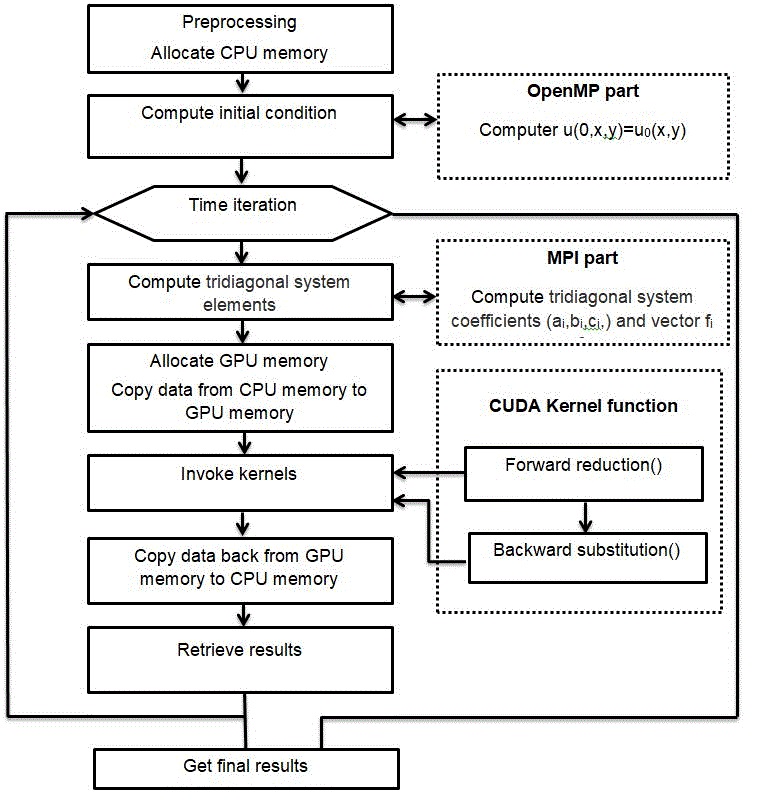}}
\end{minipage}
\caption{Flowchart of hybrid approach}
\label{fig1}
\end{figure}

These parallel technologies, CUDA, OpenMP and MPI can be combined to form a multi-layered hybrid structure, the premise is that the system has several CPU cores and at least one graphics processor. Under this hybrid structure (Figure \ref{fig1}), we can make better use of the advantages of another programming model.

\section{Experimental Results}
In this section we show the results obtained on a desktop computer with configuration 4352 cores GeForce RTX 2080 TI, NVIDIA GPU together with a CPU Intel Core(TM) i7-9800X, 3.80 GHz, RAM 64Gb.
Simulation parameters are configured as follows. Mesh size is uniform in both directions with $\Delta x = \Delta y=0.01 $, coefficients $c=1$ and numerical time step $\Delta t$ is 0.02, and simulation time is $T=1.0$, therefore the total numerical time step is 50.

Using the implicit sub-scheme \eqref{fr1}, the cyclic reduction \cite{Hock} method is performed in the $x$ direction, with the result that we get the grid function $u_{i,j}^{k+1/2}$. In the second fractional time step, using the sub-scheme \eqref{fr2}, the cyclic reduction method is performed in the direction of the $y$ axis, as the result we get the grid function $u_{i,j}^{k+1}$.

To present more realistic data we test four cases with large domain sizes of  $1024\times1024, 2048\times2048, 4096\times4096$ and $8192\times8192$.
In Table 1  we report the execution times in seconds for serial (CPU time), CUDA (GPU time), GPU+OpenMP, and CUDA+OpenMP+MPI implementations of the cyclic reduction method to the discrete problem \eqref{CN1}--\eqref{CN3}.

\begin{table}[htbp]
\caption{Execution Time (Seconds)}
\begin{tabular}{  c | c | c | c | c | c | c | c  }
\hline
N (mesh size) & CPU   & GPU  & GPU/OpenMP &  \multicolumn{3}{c}{GPU/OpenMP/MPI  }  \\
 &    &   &   &  2 CPU core & 4 CPU core  & 8 CPU core \\
\hline
$1024\times1024$& 48.13  & 24.104 & 24.151 & 24.432 & 23.232 & 22.61  \\
$2048\times2048$ & 189.677  & 45.033 & 45.01 & 35.133 & 33.571 & 30.261  \\
$4096\times4096$ & 755.614 & 122.24 & 59.996 & 58.797 & 54.223 & 51.413  \\
$8192\times8192$ & 3272.305 & 435.854 & 239.556 & 173.45 & 168.876 & 159.501  \\
\hline
\end{tabular} \\
\end{table}

\section{Conclusion and Future Work}
In this paper, we proposed the numerical solution of the 2D acoustic wave equation based on an implicit finite difference scheme using the cyclic reduction method.
And, we constructed a heterogeneous hybrid programming environment for a single PC by combining the message passing interface MPI, OpenMP, and CUDA programming.
Also, implemented parallelization of the cyclic reduction method on the graphic processing unit.
Finally, we showed how we accelerated the cyclic reduction method on the NVIDIA GPU.
From the test results of table 1  it can be seen that the acceleration method proposed by us gives a good result. Our hybrid CUDA/OpenMP/MPI implementation obtained up to 2.75 times faster results than CUDA implementation.

In future work, we are planning to improve and adapt our hybrid approach for GPU cluster and test it.

\end{document}